\documentclass[aps, prb, reprint, final, footinbib, superscriptaddress, showpacs, citeautoscript]{revtex4-1}
\usepackage[T1]{fontenc}
\usepackage[utf8]{inputenc}
\usepackage[english]{babel}
\usepackage{cmap}
\usepackage{textcomp}
\usepackage{amsmath}
\usepackage{amssymb}
\usepackage{graphicx}
\usepackage[colorlinks,allcolors=blue]{hyperref}
\usepackage[figure,figure*]{hypcap}
\usepackage{times}
\usepackage{mathptmx}

\begin{document}

\title{Polariton transport in one-dimensional channels}
\author{M.\,Yu.~Petrov}
\email{m.petrov@spbu.ru}
\affiliation{Spin Optics Laboratory, Saint-Petersburg State University,
Petrodvorets, 198504 St.~Petersburg, Russia}
\author{A.\,V.~Kavokin}
\affiliation{Spin Optics Laboratory, Saint-Petersburg State University,
Petrodvorets, 198504 St.~Petersburg, Russia} 
\affiliation{Physics and
Astronomy School, University of Southampton, Highfield, Southampton SO17
1BJ, UK}

\begin{abstract}
We study theoretically the transport of linearly polarized
exciton-polaritons in a quasi one-dimensional microcavity channel separating
two polariton condensates generated by optical pumping. The direction and
value of mass and spin currents are controlled by the relative phase and
polarisation of two condensates, as in the stationary Josephson effect.
However, due to dissipation and particle-particle interactions, the current
denisty is inhomogeneous: it strongly depends on the coordinate along the
axis of the channel. A stationary spin domain can be created in the channel,
its position would be sensitive to the phase difference between two
bordering condensates.
\end{abstract}

\maketitle

\section*{Introduction}

Exciton-polaritons are electrically neutral bosonic spin carriers. They
offer a valuable alternative to the traditional spintronics based on the
charged fermionic spin carriers, electrons and holes~\cite{ShelykhPRB04,
ShelykhSScT10, CCRevModPhys13}. Recent experiments demostrated a high
potentiality of the exciton-polaritons for the ballistic transport over
macroscopic distancies~\cite{LeyderNatPhys07, WertzNatPhys10}. Spin
Josephson~\cite{PavlovichPRB13}, optical spin Hall~\cite{KavokinPRL05},
optical Aharonov-Bohm~\cite{ShelykhPRL09} effects based on the
exciton-polaritons are widely discussed. The spin currents carried by the
exciton-polaritons are being studied experimentally by the
polarisation-resolved micro-photoluminescence~\cite{AmoNatPhoton10,
AdradosPRL11}. Meanwhile the theory of bosonic spin transport in dissipative
media is far from being built. The concept of bosonic spin conductivity is
not clearly established yet. The interplay between non-linear amplification
effects and dissipation induced by the radiative decay of exciton-polaritons
makes building up of the theory of the polariton spin currents an
interesting and non-trivial task.

This work is aimed at theoretical description of the bosonic transport in a
dissipative environment by considering a simplest model system, a
one-dimensional channel. Such a system can be realized in a semiconductor
microcavity with embedded quantum wells, in the regime of the strong
exciton-photon coupling~\cite{Microcavities}. We shall assume that the
temperature is low enough to allow for formation of a spatially coherent
polariton condensate~\cite{KasprzakNat06} able to expand over the
macroscopic distances without significant dephasing, as it has been
experimentally observed in Ref.~\onlinecite{AdradosPRL11}.

A one dimensional channel similar to those studied in fermionic systems~\cite%
{ThorntonPRL86} can be formed by the etching of the upper Bragg mirror of a
microcavity in order to create a lateral confinement for exciton-polaritons,
which would be free to move along the axis of the channel~\cite%
{WertzNatPhys10,GaoPRB12}. In contrast to the system studied in Ref.~%
\onlinecite{WertzNatPhys10}, where the channel has been closed on both
sides, we consider an open channel connecting two semi-infinite microcavity
areas, in which the polariton condensates are created by quasi-resonant
optical pumping [see the scheme in Fig.~\ref{fig:Scheme}(a)]. In this
configuration, the chemical potentials of the polariton condensates to the
left and right sides of the channel are set by the energies of two pumping
beams [see Fig.~\ref{fig:Scheme}(b)]. Moreover, the polarisation and the
phase of the both condensates are also controlled by the pumping beams (see
e.g. Refs.~\onlinecite{GippiusPRL07, LiewPRL08, SarkarPRL10, ParaisoNatMat10}%
).

We shall study the spin transfer through the channel separating two
condensates as a function of their chemical potentials, phases and
polarisations. All these parameters can be efficiently coltrolled by the
quasi-resonant pumping. We shall account for the spin-dependent
polariton-polariton interactions and the radiative decay of the polaritons
everywhere in the structure, including the channel.

The paper is organised as follows. In the Section~\ref{sec:Model} we
formulate a basic concept of the model based on the mean-field approach.
Section~\ref{sec:Results} presents the results of numerical calculations of
polariton currents in the channel at different boundary conditions. In the
last Section we conclude on the specificity of bosonic spin transport in
dissipative systems.

\section{The mean-field model\newline
}

\label{sec:Model}

\begin{figure}[t]
\includegraphics{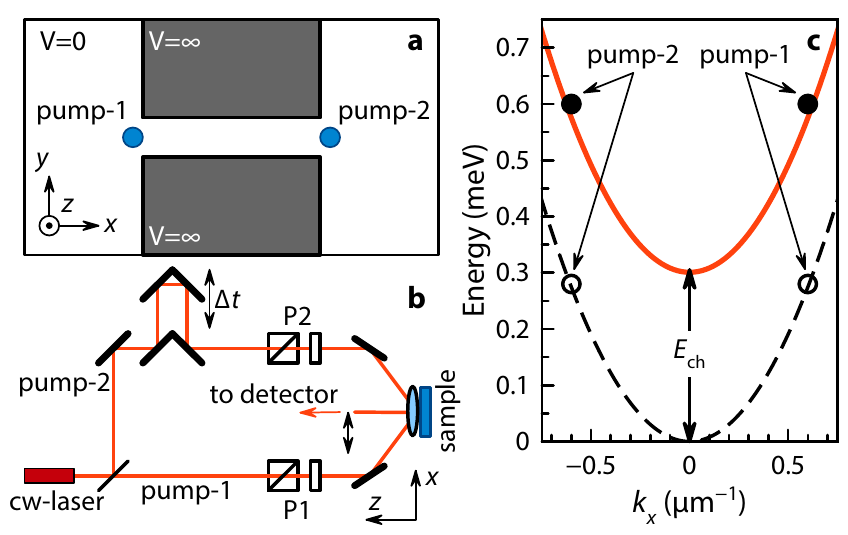}
\caption{(Color online) (a) A sketch showing the narrow channel defined by
the spatial variation of the external potential for exciton-polaritons in a
planar microcavity, $V=0$ in the channel and $V=\infty $ in the barrier
regions. Two pumping beams excite polaritons in the semi-infinite open areas
near the channel ends. (b)~The scheme of a model experiment, which implies
the continuous-wave laser excitation splitted into two pumping beams passing
through the delay line $\Delta t$ and polarizers $\mathrm{P1}$ and $\mathrm{%
P2}$. The detection system allows for collecting the spatially resolved
photoluminescence signal from the channel. (c)~The dispersion of the
lower-polariton branch in semi-infinite open areas (dashed line) and in the
channel (solid line). The pumping energy $E_{p}$ is tuned slightly above the
lower-polariton branch energy (see Fig.~\protect\ref{fig:Density}).}%
\label{fig:Scheme}
\end{figure}

In microcavities based on zinc-blend semiconductors like GaAs,
exciton-polaritons formed by heavy-hole excitons have two allowed spin
projections to the structure axis: $\sigma =\pm 1$. In a coherent system the
state of the polariton condensate can be described by a two-component order
parameter $\Psi (\mathbf{r},t)$~\cite{GippiusPRL07}: $\Psi (\mathbf{r}%
,t)=[\psi _{+1}(\mathbf{r},t);\psi _{-1}(\mathbf{r},t)]$ where $\psi
_{\sigma }(\mathbf{r},t)$ is the many body wave~function of the condensate
of polaritons with a spin projection $\sigma $ at a position $\mathbf{r}$
and a time $t$. The dynamics of $\Psi (\mathbf{r},t)$ is described by the
Gross-Pitaevskii (GP) equation~\cite{GippiusPRL07, LiewPRL08, CCPRL04,
BerloffPRL05, WoutersPRL07, MalpuechPRL07, ShelykhPRB08, AmoNatPhys09, 
WoutersPRB10, FlayacPRB11}. The stationary state for the two-component 
wave-function satisfies 
\begin{subequations}
\label{eq:GP}
\begin{align}
\hslash \omega _{p}\psi _{+}(\mathbf{r})& =\bigl[\hat{E}_{\mathrm{LP}%
}(-i\nabla )+\alpha _{1}|\psi _{+}(\mathbf{r})|^{2}+\alpha _{2}|\psi _{-}(%
\mathbf{r})|^{2}+  \notag \\
& V(\mathbf{r})-\tfrac{i\hslash }{2\tau }\bigr]\psi _{+}(\mathbf{r})+F_{p+}(%
\mathbf{r}).  \label{eq:GP1} \\
\hslash \omega _{p}\psi _{-}(\mathbf{r})& =\bigl[\hat{E}_{\mathrm{LP}%
}(-i\nabla )+\alpha _{1}|\psi _{-}(\mathbf{r})|^{2}+\alpha _{2}|\psi _{+}(%
\mathbf{r})|^{2}+  \notag \\
& V(\mathbf{r})-\tfrac{i\hslash }{2\tau }\bigr]\psi _{-}(\mathbf{r})+F_{p-}(%
\mathbf{r}).  \label{eq:GP2}
\end{align}%
Here $\hat{E}_{\mathrm{LP}}(\mathbf{k})$ is the kinetic energy operator for
the lower polariton (LP) branch, $V(\mathbf{r})$ is the position dependent
external potential, $\tau $ is the polariton lifetime, $\alpha _{1}$ and $%
\alpha _{2}$ are the polariton-polariton interaction contacts in parallel
and antiparallel spin configurations, respectively, and $F_{p\sigma }(%
\mathbf{r})$ describes the pumping of polaritons with spin projection $%
\sigma $. The term $\hslash \omega _{p}$ in the left hand side of Eqs.~%
\eqref{eq:GP} sets the chemical potential of the polaritons created by
pumping (we shall assume equal energies of left and right pumps).

The pumping terms which describe the quasi-resonant optical injection of
polaritons at the lower polariton branch at the left and right ends of the
channel are introduced following~\cite{CCPRL04}: 
\end{subequations}
\begin{align}
F_{p\sigma }(\mathbf{r},t)& =A_{\sigma }\bigl(\exp [-(\mathbf{r}-\mathbf{r}%
_{L})^{2}/\delta ^{2}+i\mathbf{k}_{p}^{L}\cdot \mathbf{r}]+  \notag \\
& \exp [-(\mathbf{r}-\mathbf{r}_{R})^{2}/\delta ^{2}-i\mathbf{k}%
_{p}^{R}\cdot \mathbf{r}]\bigr)e^{-i\omega _{p}t}.
\end{align}%
Here $\mathbf{r}_{L}=-\mathbf{r}_{R}$ and $\mathbf{k}_{p}^{L}=-\mathbf{k}%
_{p}^{R}$ are the positions and the wave vectors of the incident light
field, $\delta $ is the size of the laser-pumping spot, $A_{\sigma }$ is the
amplitude of $\sigma $~component of the pumping field, and $\omega _{p}$ is
the pumping frequency tuned so that $E_{p}=\hslash \omega _{p}$ is slightly
above the lower-polariton branch energy [see Fig.~\ref{fig:Scheme}(c)]. Due
to the lateral confinement potential $V(\mathbf{r})$ in Eqs.~\eqref{eq:GP},
the polariton dispersion in the channel is blueshifted by $E_{\mathrm{ch}}$
with respect to the dispersion of polaritons in the semi-infinite areas to
the left and right from the channel.

We solve the coupled stationary GP equations \eqref{eq:GP1} and %
\eqref{eq:GP2} using the Newton-type iteration procedure based on the
spatial discretization with the finite element method~\cite{ComsolComment}.
The following parameters are used in the calculations. The channel width is $%
D_{\mathrm{ch}}=5$~\textmu{}m and its length is $L_{\mathrm{ch}}=30$~%
\textmu{}m. The lower polariton dispersion is approximated by a parabola
characterised by polariton effective mass $m_{\mathrm{LP}}=5\cdot
10^{-5}m_{0}$ where $m_{0}$ is the mass of a free electron. The polariton
lifetime is taken either $\tau =30$~ps which is close to the values achieved
experimentally in high Q-factor cavities, or $300$~ps which is unrealistic
in the existing semiconductor microcavities, for the sake of comparison.
Polariton-polariton interaction constants $\alpha _{1}=2\cdot 10^{-10}$~meV$%
\cdot $\thinspace cm$^{2}$ and $\alpha _{2}=-0.1\alpha _{1}$ are taken from
Ref.~\onlinecite{VladimirovaPRB09}. The coordinates of pumping spots are
chosen as $y=0$ in $x=\pm 16.5$~\textmu{}m. The spot size is $\delta =0.25D_{%
\mathrm{ch}}$ and the wavevectors are $k_{x}=\pm 0.6$~\textmu{}m$^{-1}$. The
pumping energy $E_{p}$ is tuned above the bare polariton branch by $0.015$%
~meV, and the amplitude $A_{\sigma }$ is a variable parameter.

\section{Results and Discussion}

\label{sec:Results}

\begin{figure}[t]
\includegraphics{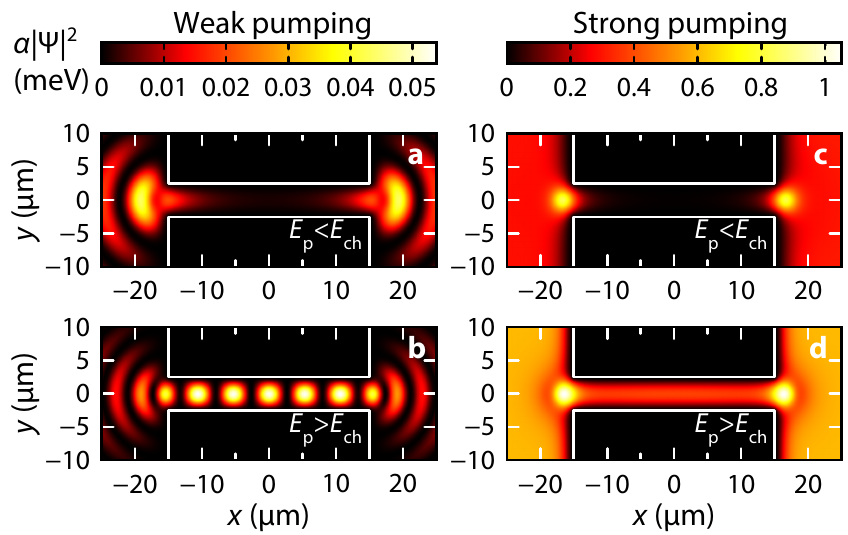}
\caption{(Color online) The calculated polariton density distributions in
the channel in the regimes of weak pumping (panels (a) and (b)) and strong
pumping (panels (c) and (d)). The energy of the pumping beams is tuned below $%
E_{\mathrm{ch}}$ (panels (a) and (c)) and above $E_{\mathrm{ch}}$ (panels
(b) and (d)).}
\label{fig:Density}
\end{figure}

We shall limit our consideration to the case of equal chemical potentials of
the condensates formed at the left and right ends of the channel. We
consider linearly polarised pumping beams so that $A_{+1}=A_{-1}=A$. No spin
is injected to the system by the pumping lasers in this case. Let us first
discuss the effect of the pumping energy $E_{p}$ and the pumping amplitude $%
A_{\sigma }$ on the polariton density profile in the channel. At low
pumping, the non-linear effects due to polariton-polariton interactions are
negligible. In this case, the classical interference effect leads to the
formation of a stationary wave with a sinusoidal profile in the channel [see
Figs.~\ref{fig:Density}(a) and \ref{fig:Density}(b)]. Once $A$ increases so
that the nonlinear terms in Eqs.~\eqref{eq:GP} start dominating over
radiative losses, the interference pattern disappears and the polariton
density in the channel becomes constant dependent on the chemical potential
set by the two pumps [see Figs.~\ref{fig:Density}(c) and \ref{fig:Density}%
(d)]. The channel may be open or closed for the polaritons if their chemical
potential is either larger or smaller than $E_{\mathrm{ch}}$, respectively
(compare panels (a) vs (b) and (c) vs (d) in Fig.~\ref{fig:Density}). The
most interesting and unusual bosonic spin transport phenomena are expected
in the regime where $E_{p}>E_{\mathrm{ch}}$ and the pumping power is
sufficiently large. This is the regime we are going to consider in the rest
of this paper.

\begin{figure*}[t]
\includegraphics{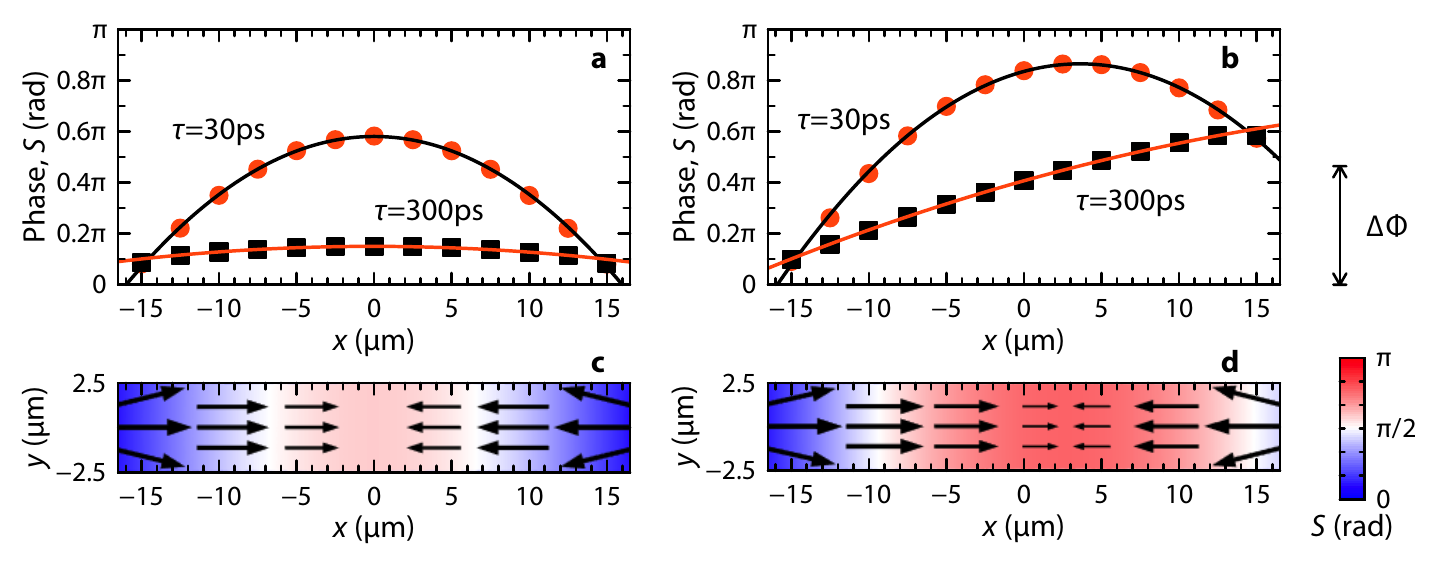}
\caption{(Color online) (a) and (b) show the phase of the condensate in the
channel $S(x)$ calculated for the different polariton lifetimes and for the
different phase shifts between pumping beams (symbols). Thin lines show the
fit of numerical results with Eq.~\eqref{eq_Phase}. (c) and (d)~show the
profiles of the phase of the condensate in the channel calculated for $%
\protect\tau =30$~ps and the same $\Delta \Phi $ as in panels (a) and (b),
respectively. Arrows show the direction and value (in log scale) of the
polariton current density.}
\label{fig:Phase}
\end{figure*}

Let us consider now the spatial behaviour of the phase of the polariton
condensate in the channel. According to the conventional definition~\cite%
{BECBook}, its order parameter can be represented as $\psi _{\sigma }(%
\mathbf{r})=\sqrt{n_{\sigma }(\mathbf{r})}e^{iS_{\sigma }(\mathbf{r})}$,
where $n(\mathbf{r})=[n_{+}(\mathbf{r});n_{-}(\mathbf{r})]$ and $S(\mathbf{r}%
)=[S_{+}(\mathbf{r});S_{-}(\mathbf{r})]$ are two-component functions, which
characterize the polariton density and phase, respectively. As shown above,
the density profile of the ground state is almost constant along the channel
axis [Fig.~\ref{fig:Density}(d)]. However, due to the dissipation, the phase
of the wave~function can vary as a function of $x$. In general, it has a
nearly parabolic shape as Figs.~\ref{fig:Phase}(a) and \ref{fig:Phase}(b)
show. The spatial variation of the phase can be found analytically by
substituting the parabolic ansatz into the GP equation and separating its
real and imaginary parts (see Appendix~\ref{sec:Appendix} for details). The
coordinate dependence of the phase writes 
\begin{equation}
S(x)=-\frac{m_{\mathrm{LP}}}{6\hslash \tau }x^{2}+\frac{\Delta \Phi }{L_{%
\mathrm{ch}}}x+S_{0}.  \label{eq_Phase}
\end{equation}%
Here $\Delta \Phi $ is the phase difference between two pumping beams governed
by the time delay $\Delta t$ [see Fig.~\ref{fig:Scheme}(b)].

\begin{figure}[b]
\includegraphics{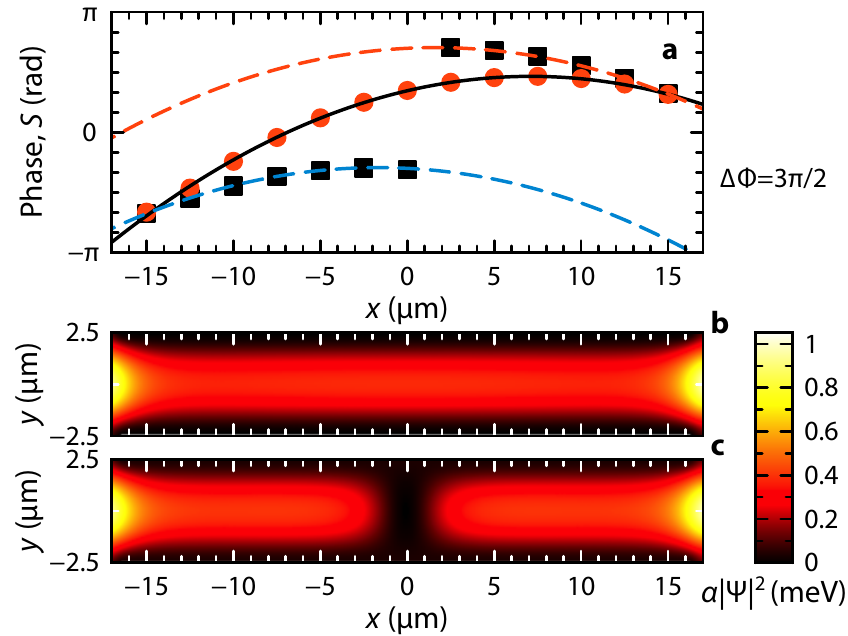}
\caption{(Color online) (a) shows the polariton phase profiles along the
axis of the channel calculated at $\Delta \Phi =3\protect\pi /2$ (symbols)
and their parabolic fits with Eq.~\eqref{eq_Phase} (lines). (b) and (c)\
show the corresponding density profiles of the components of the condensate.
The colorbar shows the scale of the corresponding blueshift.}
\label{fig:Phase2}
\end{figure}

As the phase profile of the order parameter is a non-linear function of the
coordinate along the axis of the channel, the current density defined as~%
\cite{BECBook} 
\begin{equation}
\mathbf{j}(\mathbf{r})=\frac{\hslash }{m_{\mathrm{LP}}}n(\mathbf{r})\nabla S(%
\mathbf{r})
\end{equation}%
is not conserved along the channel. As $S$ is a parabolic function of the
coordinate, the current density is a linear function of the coordinate. Zero
current point is in the center of the channel if the phases of two
condensates coincide. It can be shifted from the center to the left or right
side by changing the phase difference, $\Delta \Phi $, between two
condensates. 

\begin{figure*}[t]
\includegraphics{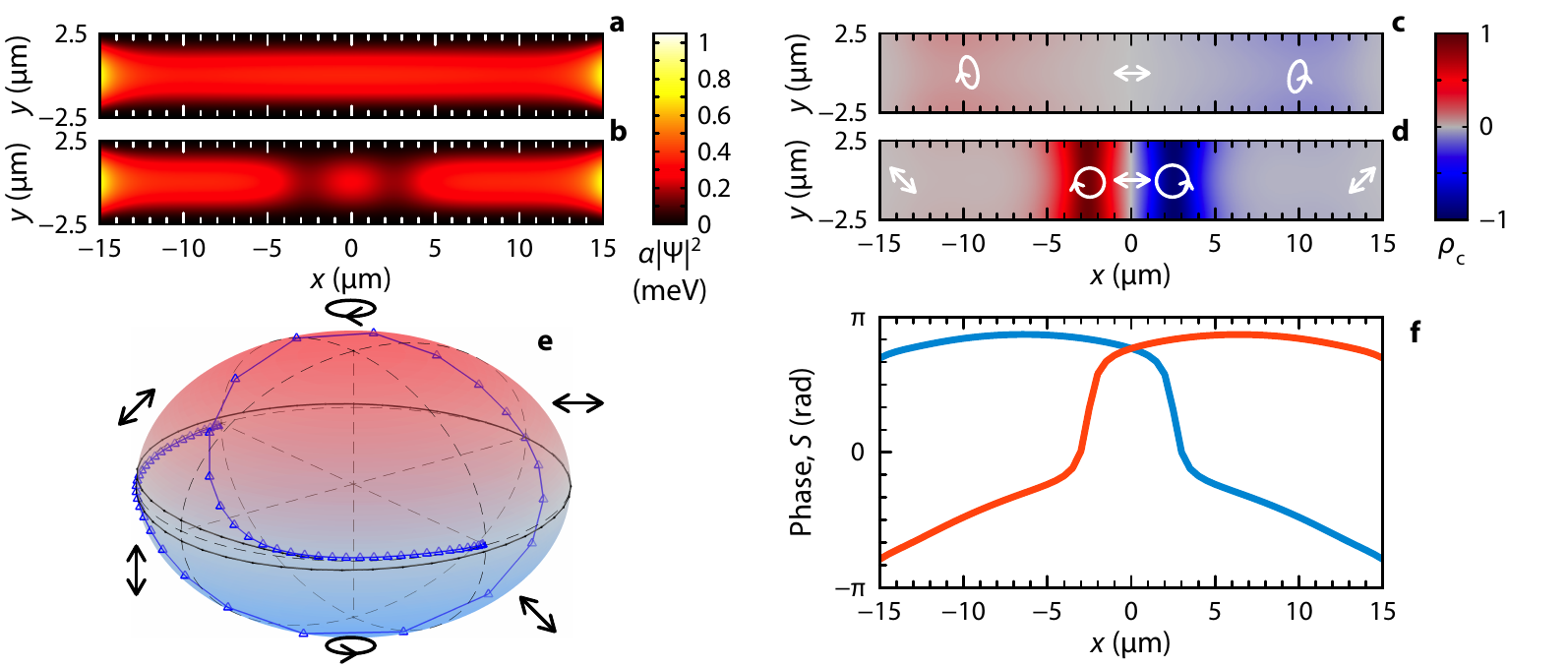}
\caption{(Color online) Polariton density distribution in the channel
calculated for zero phase shift between right and left pumping beams  $\Delta
\Phi =0$ (a) and for $\Delta \Phi =\protect\pi $ (b). (c) and (d)~show the
profiles of the circular polarization degree corresponding to the density
distributions shown in panels (a) and (b). Arrows indicate orientations of
the Stokes vector of the condensate at different points along the axis of
the channel at $y=0$. (e) shows the evolution of the Stokes vector on a
Poincar\'{e} sphere as one goes along the axis of the channel between its
left and right ends. The dotted (black) line corresponds to $\Delta \Phi =0$
and the triangled (blue) line corresponds to $\Delta \Phi =\protect\pi $.
(f) shows the phase profile of the order parameter components calculated
along the channel axis at $y=0$ for $\Delta \Phi =\protect\pi $.}%
\label{fig:Polarizations}
\end{figure*}

The most interesting effect occurs if the phase shift between the left and
right condensates is large, i.e. $|\Delta \Phi |>L_{\mathrm{ch}}^{2}m_{%
\mathrm{LP}}/(12\hslash \tau )$. The polariton flows become unstable: two
solutions of the GP equations \eqref{eq:GP} have been found in this case
[see Fig.~\ref{fig:Phase2}(a)]. The first one [circles in Fig.~\ref%
{fig:Phase2}(a)] corresponds to a strong variation of the density profile as
it is shown in Fig.~\ref{fig:Phase2}(b). The second one [squares in Fig.~\ref%
{fig:Phase2}(a)] allows for a weaker spatial variation of the phase of the
condensate. In order to maintain the phase difference $\Delta \Phi $ between
two ends of the channel, the dependence $S(x)$ bounces by $\pi $ in the
middle point. The phase jump corresponds to the break in the condensate
profile, which is seen in Fig.~\ref{fig:Phase2}(c). The position of the
\textquotedblleft hole\textquotedblright\ in the polariton density of a
given spin component is sensitive to $\Delta \Phi $.

This fragmentation effect manifests the spontaneous pattern formation in a
non-linear bosonic system. Similar effects have been observed in polariton
condensates excited in planar cavities by non-resonant optical pumping with
an elliptically polarised light~\cite{ManniPRL11}. Manni \emph{et al.} (Ref.~%
\onlinecite{ManniPRL11}) observed a single-energy condensed state featuring
a denisty and polarisation pattern detectable in the polarisation-resolved
photoluminescence. Also, in the microcavity stripes~\cite{WertzNatPhys10}
the standing-wave states have been observed, which manifest themselves by
patterns in near-field photoluminescence spectra.

In our case, however, there is no standing wave along the channel. The
effect predicted here is a consequence of the destructive interference of
two polariton fluids coming from the opposite ends of the channel, in a
non-linear regime.

Further non-trivial effects may be observed if the polarisations of two
pumps do not coincide. As an example, we consider pumping with two linearly
polarised laser beams having orthogonal polarisation planes. Figs.~\ref%
{fig:Polarizations}(a)--\ref{fig:Polarizations}(d) demonstrate the density
profile of the condensate and the corresponding spatial distribution of the
circular polarisation degree. The circular polarisation degree of the
condensate is defined as 
\begin{equation}
\rho _{c}=\frac{|\psi _{+}|^{2}-|\psi _{-}|^{2}}{|\psi _{+}|^{2}+|\psi
_{-}|^{2}}.
\end{equation}%
One can see that the density profile of the order parameter remains uniform
along the channel. However, due to the interference of two linearly
polarised waves, a domain of circular polarisation is formed in the middle
of the channel. In the non-linear regime, this effect is strongly modified
due to\ attraction of polaritons having opposite spins, which is accounted
for in the GP equations by the terms proportional to $\alpha _{2}$. The
polarisation and spin distribution in the channel are described by variation
of the Stokes vector of the polariton condensate along the axis of the
channel. The trajectory of the Stokes vector on a Poincar\'{e} sphere
calculated along the $x$~axis ($y=0$) is shown in Fig.~\ref%
{fig:Polarizations}(e).

If the pumping beams have exactly opposite phases, the order parameter
patterning appears right in the center of the channel. A bright spot appears
between two dark areas [Fig.~\ref{fig:Polarizations}(b)]. This effect is
very clearly seen in the circular polarization degree profile [Fig.~\ref%
{fig:Polarizations}(d)]. The stationary domains of left- and right-circular
polarizations are clearly seen in this regime. The domains may be shifted
along the axis of the channel by changing the phase difference between two
bordering condensates.

The circular polarization domains (spin domains) are formed if the
fragmentation of the polariton density occurs at the locations placed
symmetrically around $x=0$. The break in polariton density for one of the
spin components is accompanied by the phase jump by $\pi $ as it is shown in
Fig.~\ref{fig:Polarizations}(d). In the same time, in the opposite
polarization the signal varies smoothly. The appearance of circular
polarisation domains manisfests separation of phases in the spinor bosonic
system. This effect may be seen as a topological transition in a driven and
dissipative non-linear system. Experimentally, the spin domains may be
detected by near-field photoluminescence. The positions of domain walls are
strongly sensitive to the phase shift $\Delta \Phi $. In particular, the
deviation of $\Delta \Phi $ from $\pm \pi $ points by $\pi /6$ shifts the
circular domains by a half of their size. Increasing further the phase
difference between two pumps one destroys the domain structure.

\section{Summary}

\label{sec:Summary}

To summarize, we have studied theoretically the polariton mass and spin
transport in a one-dimensional microcavity channel. We have shown that in
the stationary regime, under continuous wave quasi-resonant excitation by
two cw pumping beams, the polariton current in the channel is sensitive to the
energy and intensity, relative phase and polarisation of the pumping beams.
In the nonlinear regime, we have found that the value and direction of
polariton currents varies along the axis of the channel. The polariton flow
is found to be very sensitive to the phase difference between pumping beams.
At the phase differences between two pumps close to $\pm \pi $ the
fragmentation of the condensate takes place. If two pumps have orthogonal
linear polarizations the spin domains appear.

\acknowledgements
We thank K.V. Kavokin and I.V. Ignatiev for useful discussions. MYP
acknowledges support of Computer Center of Saint-Petersburg State University
providing with the software. The financial support from the Russian Ministry
of Education and Science (contract No. 11.G34.31.0067) is acknowledged.

\begin{appendix}

\section{Derivation of Eq.~\eqref{eq_Phase}}
\label{sec:Appendix}

To derive the coordinate dependence of the phase profile of the polariton wave function 
given by Eq.~\eqref{eq_Phase} one can use a simplified 1D model. In this case, 
by applying the Madelung transformation of the wave~function $\psi_\pm = \sqrt{n_\pm}%
e^{iS_\pm}$ and by substituting it into the GP equations \eqref{eq:GP} one easily 
obtains a system of two interconnected differential equations for real and imaginary parts of the order parameter. 
The imaginary part is described by a continuity equation
\begin{equation}
-\frac{\hslash}{m} \left( \nabla n_\pm \nabla S_\pm + n_\pm \nabla^2 S_\pm \right) = 
\frac{n_\pm}{\tau}.\label{eq_imagGP}
\end{equation} 
Assuming that the density of the polaritons smoothly varies along the channel, we neglect 
the quantum pressure term (Thomas-Fermi limit) in the real part of the GP equation. 
It writes
\begin{equation}
\frac{\hslash^2}{2m} (\nabla S)^2 + \alpha_1 n_\pm + \alpha_2 n_\mp = \hslash \omega_p.
\label{eq_realGP}
\end{equation}
As we are interested here only in the dependence of functions $n_\pm$ and $S_\pm$ 
on $x$~coordinate we substitute them by parabolic functions
\begin{subequations}
\begin{align}
S_\pm = S_2^\pm x^2 + S_1^\pm x + S_0^\pm,\label{eq_parabolic_S}\\
n_\pm = n_2^\pm x^2 + n_1^\pm x + n_0^\pm.\label{eq_parabolic_n}
\end{align}
\end{subequations}
Additionally, the pumping defines the boundary condition
\begin{equation}
S_\pm\Bigl(\frac{L_{\mathrm{ch}}}{2}\Bigr) - S_\pm\Bigl(-\frac{L_{\mathrm{ch}}}{2}\Bigr) = 
\Delta\Phi,
\end{equation}
from which $S_1^\pm = \Delta\Phi/L_{\mathrm{ch}}$. The remaining coefficients can be found 
by substituting Eqs.~\eqref{eq_parabolic_S} and \eqref{eq_parabolic_n} into 
Eqs.~\eqref{eq_imagGP} and \eqref{eq_realGP} and by equating the factors at equal powers.

\end{appendix}


\begin{thebibliography}{99}
\bibitem{ShelykhPRB04} I. A. Shelykh, K. V. Kavokin, A. V. Kavokin, G.
Malpuech, P. Bigenwald, H. Deng, G. Weihs, and Y. Yamamoto, Phys. Rev. B 
\textbf{70}, 035320 (2004).

\bibitem{ShelykhSScT10} I. A. Shelykh, A. V. Kavokin, Y. G. Rubo, T. C. H.
Liew, and G. Malpuech, Semicond. Sci. Technol. \textbf{25}, 013001 (2010).

\bibitem{CCRevModPhys13} I. Carusotto and C. Ciuti, Rev. Mod. Phys. \textbf{%
85}, 299 (2013).

\bibitem{LeyderNatPhys07} C. Leyder, M. Romanelli, J. P. Karr, E. Giacobino,
T. C. H. Liew, M. M. Glazov, A. V. Kavokin, G. Malpuech, and A. Bramati,
Nat. Phys. \textbf{3}, 628 (2007).

\bibitem{WertzNatPhys10} E. Wertz, L. Ferrier, D. D. Solnyshkov, R. Johne,
D. Sanvitto, A. Lema\^{\i}tre, I. Sagnes, R. Grousson, A. V. Kavokin, P.
Senellart, G. Malpuech, and J. Bloch, Nat. Phys. \textbf{6}, 860 (2010).


\bibitem{PavlovichPRB13} G. Pavlovic, G. Malpuech, and I. A. Shelykh, Phys.
Rev. B \textbf{87}, 125307 (2013).

\bibitem{KavokinPRL05} A. Kavokin, G. Malpuech, and M. Glazov, Phys. Rev.
Lett. \textbf{95}, 136601 (2005).

\bibitem{ShelykhPRL09} I. A. Shelykh, G. Pavlovic, D. D. Solnyshkov, and G.
Malpuech, Phys. Rev. Lett. \textbf{102}, 046407 (2009).

\bibitem{AmoNatPhoton10} A. Amo, T.C.H. Liew, C. Adrados, R. Houdre, E.
Giacobino, A. V. Kavokin and A. Bramati, Nat. Photon. \textbf{4}, 361 (2010).

\bibitem{AdradosPRL11} C. Adrados, T. C. H. Liew, A. Amo, M. D. Mart\'{\i}n,
D. Sanvitto, C. Ant\'{o}n, E. Giacobino, A. Kavokin, A. Bramati, and L. Vi%
\~{n}a, Phys. Rev. Lett. \textbf{107}, 146402 (2011).

\bibitem{Microcavities} A. V. Kavokin, J. J. Baumberg, G. Malpuech, and F.
P. Laussy, \emph{Microcavities} (Oxford University, New York, 2007).

\bibitem{KasprzakNat06} J. Kasprzak, M. Richard, S. Kundermann, A. Baas, P.
Jeambrun, J. M. J. Keeling, F. M. Marchetti, M. H. Szyma\'nska, R. Andr\'e,
J. L. Staehli, V. Savona, P. B. Littlewood, B. Deveaud, and Le Si Dang,
Nature (London) \textbf{443}, 409 (2006).

\bibitem{ThorntonPRL86} T. J. Thornton, M. Pepper, H. Ahmed, D. Andrews, and
G. J. Davies, Phys. Rev. Lett. \textbf{56}, 1198 (1986).

\bibitem{GaoPRB12} T. Gao, P. S. Eldridge, T. C. H. Liew, S. I. Tsintzos, G.
Stavrinidis, G. Deligeorgis, Z. Hatzopoulos, and P. G. Savvidis, Phys. Rev.
B \textbf{85}, 235102 (2012).

\bibitem{GippiusPRL07} N. A. Gippius, I. A. Shelykh, D. D. Solnyshkov, S. S.
Gavrilov, Y. G. Rubo, A. V. Kavokin, S. G. Tikhodeev, and G. Malpuech, Phys.
Rev. Lett. \textbf{98}, 236401 (2007).

\bibitem{LiewPRL08} T. C. H. Liew, A. V. Kavokin, and I. A. Shelykh, Phys.
Rev. Lett. \textbf{101}, 016402 (2008).

\bibitem{SarkarPRL10} D. Sarkar, S. S. Gavrilov, M. Sich, J. H. Quilter, R.
A. Bradley, N. A. Gippius, K. Guda,V. D. Kulakovskii, M. S. Skolnick, and D.
N. Krizhanovskii, Phys. Rev. Lett., \textbf{105}, 216402 (2010).

\bibitem{ParaisoNatMat10} T. K. Para\"iso, M.Wouters, Y. L\'eger, F.
Morier-Genoud and B. Deveaud-Pl\'edran, Nat. Mat. \textbf{9}, 655 (2010).

\bibitem{CCPRL04} I. Carusotto and C. Ciuti, Phys. Rev. Lett. \textbf{93},
166401 (2004).

\bibitem{BerloffPRL05} N. G. Berloff, Phys. Rev. Lett. \textbf{94}, 010403
(2005).

\bibitem{WoutersPRL07} M. Wouters and I. Carusotto, Phys. Rev. Lett. 
\textbf{99}, 140402 (2007).

\bibitem{MalpuechPRL07} G. Malpuech, D. D. Solnyshkov, H. Ouerdane, M. M.
Glazov, and I. Shelykh, Phys. Rev. Lett. \textbf{98}, 206402 (2007).

\bibitem{ShelykhPRB08} I. A. Shelykh, D. D. Solnyshkov, G. Pavlovic, and G.
Malpuech, Phys. Rev. B \textbf{78}, 041302(R) (2008).

\bibitem{AmoNatPhys09} A. Amo, J. Lefr\`ere, S. Pigeon, C. Adrados, 
C. Ciuti, I. Carusotto, R. Houdr\'e, E. Giacobino, and A. Bramati, 
Nat. Phys. \textbf{5}, 805 (2009).

\bibitem{WoutersPRB10} M. Wouters, T. C. H. Liew, and V. Savona, Phys. Rev.
B \textbf{82}, 245315 (2010).

\bibitem{FlayacPRB11} H. Flayac, D. D. Solnyshkov, and G. Malpuech, Phys.
Rev. B \textbf{83}, 045412 (2011).

\bibitem{ComsolComment} Comsol Multiphysics v4.2 has been used for numerical
solution of coupled GP equations given by Eqs.~\eqref{eq:GP1} and \eqref{eq:GP2}. 
The second order Lagrange finite elements on the triangular mesh with $0.1$~\textmu{}m
of lateral element size have been used. Fine tuning of the nonlinear solver
allows us to achieve convergence error in the order of $10^{-4}$ to ensure
numerical stability of the solution.

\bibitem{VladimirovaPRB09} M. Vladimirova, S. Cronenberger, D. Scalbert, M.
Nawrocki, A. V. Kavokin, A. Miard, A. Lema\^itre, and J. Bloch Phys. Rev. B 
\textbf{79}, 115325 (2009).

\bibitem{BECBook} L. P. Pitaevskii and S. Stringari, \emph{Bose-Einstein
Condensation} (Clarendin, Oxford 2003).

\bibitem{ManniPRL11} F. Manni, K. G. Lagoudakis, T. C. H. Liew, R. Andr\'e,
and B. Deveaud-Pl\'edran, Phys. Rev. Lett. \textbf{107}, 106401 (2011).
\end{thebibliography}
\end{document}